\documentclass[aps,prl,twocolumn,amssymb,showpacs,nofootinbib]{revtex4}
\usepackage{graphicx}
\usepackage{amsmath}
\usepackage{amssymb}
\usepackage{bm}
\usepackage{color}

\def\be{\begin{equation}}
\def\ee{\end{equation}}
\def\ba{\begin{eqnarray}}
\def\ea{\end{eqnarray}}

\def\mpl{m_{\rm p}}
  \def\be{\begin{equation}}
\def\ee{\end{equation}}
 \def\bi{\begin{itemize}}
 \def\ei{\end{itemize}}
  \def\ben{\begin{enumerate}}
\def\een{\end{enumerate}}
  \def\bt{\begin{tabular}}
\def\et{\end{tabular}}
\def\bc{\begin{center}}
\def\ec{\end{center}}

\def\bea{\begin{eqnarray}}
\def\eea{\end{eqnarray}}

\begin{document}

\title{The Adiabatic Instability on Cosmology's Dark Side}
\author{Rachel Bean$^{1}$}
\author{\'Eanna \'E. Flanagan$^{2}$}
\author{Mark Trodden$^{3}$}

\affiliation{$^{1}$ Department of Astronomy, Cornell University, Ithaca, NY 14853, USA}
\affiliation{$^{2}$Laboratory for Elementary Particle Physics, Cornell University, Ithaca, NY 14853, USA.}
\affiliation{$^{3}$Department of Physics, Syracuse University, Syracuse, NY 13244, USA}
\date{\today}

\begin{abstract}
We consider theories with a nontrivial coupling between the
matter and dark energy sectors. We describe
a small scale instability that can occur in such models
when the coupling is strong compared to gravity, generalizing and correcting earlier
treatments. The instability is characterized by a negative sound speed squared of
an effective coupled dark matter/dark energy fluid. Our results are general, and applicable to a wide class of
coupled models and provide a powerful, redshift-dependent tool, complementary to other constraints,
with which to rule many of them out. A detailed analysis and applications to a range of
models are presented in a longer companion paper.

\end{abstract}

\maketitle

In order for our cosmological models to provide an accurate fit to current observational data, it is necessary to postulate two dramatic augmentations of the assumption of baryonic matter interacting gravitationally through Einstein's equations - dark matter and dark energy.
A logical possibility is that these dark sectors interact with each
other or with the normal matter
\cite{combined,Das:2005yj,Kesden:2006}.
A number of models have been proposed that exploit this possibility
to address, for example, the coincidence problem.

Such models face a range of existing constraints
arising from both particle physics and gravity. In this letter we consider
perturbations around the cosmological solution and demonstrate the existence of a
dynamical instability which we term the {\it adiabatic instability}.
This instability is characterized by a negative sound speed squared of the
effective coupled fluid~\cite{Kaplinghat:2006jk,Bjaelde:2007ki} and
was first discovered~\cite{Afshordi:2005ym} in a context slightly
different to that considered here - the mass varying neutrino model of
dark energy.
Our aim here is to give a general treatment of the instability, applicable to a wide
class of models, to identify the regimes in which the instability occurs, and to delineate the
resulting redshift-dependent constraints.

\medskip
\noindent
{\it Class of Models:} We begin from the following action
\bea
S[g_{ab},\phi,\Psi_{\rm j}]
&=& \int d^4x\sqrt{-g}
\left[ \frac{1}{2} \mpl^2 R
-\frac{1}{2} (\nabla \phi)^2
 - V(\phi)
\right] \nonumber \\
&&+ \Sigma_{\rm j} S_{\rm j}[e^{2 \alpha_{\rm j}(\phi)} g_{\mu\nu}, \Psi_{\rm j}] \ ,
\label{action0}
\eea
where $g_{\mu\nu}$ is the Einstein frame metric, $\phi$ is a scalar field
which acts as dark energy, and $\Psi_{\rm j}$ are the matter fields. Here we have adopted
a signature (-,+,+,+) and defined the reduced Planck mass by $\mpl^2 \equiv (8\pi G)^{-1}$.
The functions $\alpha_{\rm j}(\phi)$ are couplings to the j${}^{th}$ matter sector.
This general action encapsulates many models studied in the
literature~\cite{combined1}.
The field equations are:
\bea
\mpl^2 G_{\mu\nu} &=& \nabla_\mu \phi \nabla_\nu \phi - \frac{1}{2} g_{\mu\nu}
(\nabla \phi)^2 - V(\phi) g_{\mu\nu} \nonumber \\
&& + \sum_{\rm j} e^{4 \alpha_{\rm j}(\phi)}
\left[ ({\bar \rho}_{\rm j} + {\bar p}_{\rm j}) u_{{\rm j}\,\mu} u_{{\rm j}\,\nu} + {\bar p}_{\rm j} g_{\mu\nu} \right] \ ,
\label{ee0d}
\eea
\be
\nabla_\mu \nabla^\mu \phi - V'(\phi) = \sum_{\rm j} \alpha_{\rm j}'(\phi) e^{4 \alpha_{\rm j}(\phi)}
({\bar \rho}_{\rm j} - 3 {\bar p}_{\rm j} ) \ ,
\label{eq:scalar10a}
\ee
where we have treated the matter field(s) in the j${}^{th}$ sector as a fluid with density
${\bar \rho}_{\rm j}$ and pressure ${\bar p}_{\rm j}$ as measured in
the frame $e^{2 \alpha_{\rm j}} g_{\mu\nu}$, and with
4-velocity $u_{{\rm j}\,\mu}$ normalized according to $g^{\mu\nu} u_{{\rm j}\,\mu} u_{{\rm j}\,\nu}=-1$.

We consider models with a baryonic sector ($\alpha_b(\phi)$) and a
composite dark matter sector, with one coupled species with density
$\rho_c$ and coupling $\alpha_c(\phi) = \alpha(\phi)$, and another
uncoupled species with density $\rho_{co}$ and coupling $\alpha_{co}=0$.
Neglect the gravitational effect of the baryons, using ${\bar p}_c =
{\bar p}_{co}=0$, and defining $\rho_{\rm j} =
e^{3 \alpha_{\rm j}} {\bar \rho}_{\rm j}$ gives
\bea
\mpl^2 G_{\mu\nu} &=& \nabla_\mu \phi \nabla_\nu \phi - \frac{1}{2} g_{\mu\nu}
(\nabla \phi)^2 - V(\phi) g_{\mu\nu}\nonumber
\\ && + e^{\alpha(\phi)}\rho_c u_{c\mu} u_{c\nu}+ \rho_{co} u_{co\mu} u_{co\nu} \ ,
\label{ee}
\eea
and $
\nabla_\mu \nabla^\mu \phi - V_{\rm eff}'(\phi) = 0,
$
where we have defined an effective potential by
$
V_{\rm eff}(\phi) = V(\phi) + e^{\alpha(\phi)}\rho_c \ .
$
The fluid obeys $\nabla_\mu ( \rho_c u_c^\mu) =0$,
and $u_c^\nu \nabla_\nu u_c^\mu  = - (g^{\mu\nu} + u_c^\mu u_c^\nu) \nabla_\nu \alpha$.

\medskip
\noindent
{\it Adiabatic regime:} The effective potential $V_{\rm eff}(\phi)$ may have a minimum resulting from the
competition between the two distinct terms. If the timescale or lengthscale for $\phi$ to adjust
to the changing position of the minimum of $V_{\rm eff}$ is shorter
than that over which the background
density changes, the field $\phi$ will adiabatically track
this minimum~\cite{Das:2005yj}.
In this case the coupled CDM component together with $\phi$ together
act as a single fluid
with an effective energy density $\rho_{\rm eff}$
and effective pressure $p_{\rm eff}$:
\be
\rho_{\rm eff}(\rho_c) = e^{\alpha[\phi_{\rm m}(\rho_c)]} \rho
+ V[\phi_{\rm m}(\rho_c)] \ ,
\label{rhoeff}
\ee
\be
p_{\rm eff}(\rho_c) = - V[\phi_{\rm m}(\rho_c)] \ .
\label{peff}
\ee
Here $\phi_{\rm m}(\rho_c)$ is the solution of the algebraic
equation
\be
V_{\rm eff}'(\phi)= V'(\phi) + \alpha'(\phi) e^{\alpha(\phi)} \rho_c =0
\label{eq:alg}
\ee
for $\phi$.
Eliminating $\rho_c$ between Eqs.\ (\ref{rhoeff}) and (\ref{peff}) gives
the equation of state $p_{\rm eff} = p_{\rm eff}(\rho_{\rm eff})$.

For cosmological background solutions, we assume
that the coupled fluid acts as the
source of the cosmic acceleration.
In the adiabatic approximation, the effective fluid description is valid
for the background cosmology and for linear and nonlinear perturbations. Therefore,
the equation of state of perturbations is the same as that of the background cosmology, and
the matter and scalar field evolve as one effective fluid,
obeying the usual fluid equations of motion with the given effective
equation of state.

A necessary condition for the validity of the adiabatic approximation is that the
lengthscales or timescales ${\cal L}$ over which the density $\rho_c$ varies
are large compared to inverse of the effective mass
\be
m_{\rm eff}(\rho_c)^2 = \left. \frac{\partial^2 V_{\rm eff}}{\partial \phi^2}
  (\phi,\rho_c) \right|_{\phi = \phi_{\rm m}(\rho_c)}
\label{meffdef}
\ee
of the scalar field.  More precisely, we can show that the condition is \cite{paper2}
\be
\frac{ d \ln V[\phi_{\rm m}(\rho_c)]}{d \ln \rho_c} \left(
  \frac{1}{m_{\rm eff}^2 {\cal L}^2} \right) \ll 1;
\label{condt1}
\ee
this condition is necessary to justify dropping the terms
involving the gradient of $\phi$ from the fluid and Einstein
equations.  In most situations the logarithmic derivative factor is of
order unity and can be neglected.  In Ref.\ \cite{paper2} we also
derive a non-local sufficient condition for the validity of the
approximation, which generalizes conditions in the literature for the
chameleon (thin-shell condition) \cite{Khoury:2003aq,Khoury:2003rn} and $f(R)$ modified gravity
\cite{Sawicki:2007tf} models.
Condition (\ref{condt1}) is not very stringent; many dark energy
models admit regimes where it is satisfied for the background and for
linearized perturbations over a range of scales.

In the adiabatic regime, the
inferred dark energy equation of state parameter in the case $\alpha_b=0$
is
\be
w = \frac{-1}{1 - (1-e^{(\alpha_0 -\alpha)} ) \frac{d \ln V }{d \alpha} } \ ,
\ee
with $\alpha_0\equiv\alpha(\phi_0)$ the value today. Thus, $w$ is precisely $-1$ today,
and generically satisfies $w < -1$ in the past~\cite{Das:2005yj,paper2}.


\medskip
\noindent
{\it Adiabatic instability:} We write the
potential $V(\phi)$ as a function $V(\alpha)$ of the
coupling function $\alpha(\phi)$ by eliminating $\phi$.
This gives, from Eqs.\ (\ref{rhoeff}) and (\ref{eq:alg}),
\be
\rho_{\rm eff} = V + e^{ \alpha} \rho_c = V - \frac{
  dV/d\phi}{d\alpha/d\phi} = V - \frac{dV}{d\alpha} \ .
\ee
The
square of the adiabatic sound speed, $c_{a}^{2}=\dot{P}/\dot\rho$ is then given by
\be
\frac{1}{c_a^2} = \frac{d \rho_{\rm eff}}{dp_{\rm eff}}
= \frac{ d\rho_{\rm eff} / d\alpha}{d p_{\rm eff} / d\alpha}
= -1 + \frac{ \frac{d^2 V}{d\alpha^2} }{
  \frac{dV}{d\alpha}} \ .
\label{soundspeed}
\ee
In the adiabatic regime the effective sound speed, relating to local perturbations in pressure and density, $c_s^2(k,a)\equiv \delta P(k,a)/\delta\rho(k,a)$, tends towards the adiabatic sound speed and is {\it always negative}, since
$dV/d\alpha$ must be negative so that Eq.\ (\ref{eq:alg}) admits a solution,
and $d^2V/d\alpha^2$ must be positive so that (\ref{meffdef}) yields a
positive $m_{\rm eff}^2$. From here in, we consider the regime in which this adiabatic limit has been reached, and take $c_s^2=c_a^2$.

Consider now a perturbation with lengthscale ${\cal L}$.
In order to be in the adiabatic regime we require ${\cal L} \gg m_{\rm
 eff}^{-1}$.  The negative sound speed squared will cause an
exponential growth of the mode, as long as the
growth timescale $\sim {\cal L} / \sqrt{|c_s^2|}$ is short compared to
the local gravitational timescale $\mpl /\sqrt{\rho_{\rm eff}(\rho_c)}$.
Combining Eqs.\ (\ref{rhoeff}), (\ref{eq:alg}) and (\ref{meffdef})
yields $c_s^2 m_{\rm eff}^2 = (\alpha')^2 V_{,\alpha} = (\alpha')^2
\rho_{\rm eff} / (V/V_{,\alpha}-1)$, and therefore
the instability will operate in the range of lengthscales given by
\be
\frac{1}{m_{\rm eff}(\rho_c)} \ll  {\cal L} \ll \frac{\mpl
|\alpha^\prime[\phi_{\rm m}(\rho_c)]|}{m_{\rm eff}(\rho_c)}
\sqrt{\frac{1}{1 - \frac{1}{\frac{d\ln V}{d \alpha}}}}.
\label{range}
\ee
Here the quantity $d \ln V / d\alpha(\alpha)$
on the right hand side is
expressed as a function of $\phi$ using $\alpha = \alpha(\phi)$, and
then as a function of the density using $\phi = \phi_{\rm m}(\rho_c)$.
In order for this range of scales to  be non empty, the dimensionless
coupling $\mpl |\alpha'|$ must be large compared to unity, i.e., the
scalar mediated interaction between the dark matter particles must be
strong compared to gravity.

There are two different ways of describing and understanding the instability, depending
on whether one thinks of the
scalar-field mediated forces as ``gravitational'' or ``pressure'' forces.
In the Einstein frame, the
instability is independent of gravity, since it is present even when
the metric perturbation due to the fluid can be
neglected.  In the adiabatic regime the acceleration due the scalar
field is a gradient of a local function of the density, which can be thought of as a pressure.  The net
effect of the scalar interaction is to give a contribution to the
specific enthalpy $h(\rho_c) = \int dp/\rho_c$ of any fluid which is independent
of the composition of the fluid.
If the net sound speed squared of the
fluid is negative, then there exists an instability in accord with our
usual hydrodynamic intuition.

In the Jordan
frame description, however, the instability involves gravity.
The effective Newton's constant describing the interaction of dark matter with itself is
\be
G_{cc} = G \left[ 1
  + \frac{2 \mpl^2 \alpha^\prime(\phi)^2 }{1 +
\frac{m_{\rm eff}^2}{ {\bf k}^2 }
} \right],
\label{Gformula0}
\ee
where ${\bf k}$ is a spatial wavevector \cite{paper2}.
At long lengthscales the scalar interaction is
suppressed and $G_{cc} \approx G$.  At short lengthscales,
the scalar field is effectively massless and
$G_{cc}$ asymptotes to a constant.
However, when $\mpl |\alpha^\prime| \gg 1$ there is an intermediate range
\be
m_{\rm eff}(\mpl |\alpha^\prime|)^{-1} \ll k \ll m_{\rm eff}
\label{range3}
\ee
over which the effective Newton's constant increases like $G_{cc} \propto {\bf k}^2$.
This interaction behaves just
like a (negative) pressure in the hydrodynamic equations.
This
explains why the the effect of the scalar interaction can be thought
of as either pressure or gravity in the range of scales (\ref{range3}).
Note that the range of scales (\ref{range3}) coincides with with the
range (\ref{range}) derived above, up to a logarithmic correction factor.

From this second, Jordan-frame point of view, the instability is
simply a Jeans instability.  In a cosmological background
the CDM fractional density perturbation traditionally exhibits power-law growth on subhorizon scales because Hubble damping competes with the exponential (Jeans) instability one might expect on a timescale of $1/ \sqrt{G \rho}$.
In our case, however, the gravitational self-interaction of the mode is governed by $G_{cc}(k)$ instead of $G$, and consequently in the range (\ref{range3}) where $G_{cc} \gg G$ the timescale for the Jeans instability is much shorter than the Hubble damping time.
Therefore the Hubble damping is ineffective and the Jeans instability
causes approximate exponential growth.

\medskip
\noindent
{\it Examples of Models:} For single component dark matter models, one
can find coupled models in the adiabatic regime \cite{Das:2005yj,Sawicki:2007tf}.
However in the strong coupling limit $\mpl |\alpha'| \gg 1$ of
interest here, they typically do not yield acceptable background
cosmologies.  Therefore we focus on composite dark matter models.

As a first example we consider a constant coupling function
and an exponential potential
\be
\alpha(\phi) = -\beta C \frac{\phi}{\mpl} \ , \ \ \ V = V_0 e^{-\lambda \phi/\mpl} \ ,
\label{constantcoupling}
\ee
where $\beta\equiv \sqrt{2/3}$ and $C<0$ and $\lambda$ are constants.
The Friedmann equation in the adiabatic limit is then $3 \mpl^2 H^2 = V + e^\alpha \rho_{co} + \rho_c$, in which
the first two terms on the right hand
side act like a fluid that, for
$|C| \gg 1$, approaches a cosmological constant.  Thus, the background
cosmology is close to $\Lambda$CDM for large enough $|C|$.
Since the fraction of coupled dark matter is $\Omega_{co} = e^\alpha \rho_{co} / (3 \mpl^2 H^2)$, in the asymptotic adiabatic regime, $\Omega_{co}
=\lambda (1 - \Omega_c)/(\lambda-\beta\, C)$, and $\Omega_c \sim 0.3$ today, $\Omega_{co}$ must be small for
large coupling, $|C| \gg 1$.
If the parameters of the model are chosen so that $\Omega_c \sim 1$
today, then the maximum and minimum lengthscales for the instability are
${\cal L}_{\rm max} \sim H_0^{-1}$ and ${\cal L}_{\rm min} \sim
(H_0 \beta |C|)^{-1}$.
Taking the Jeans view, it is then possible to show~\cite{paper2} that
the instability should operate whenever modes are inside the horizon
and in this range.

\begin{figure}[t]
\begin{center}
\includegraphics[width=3.5in]{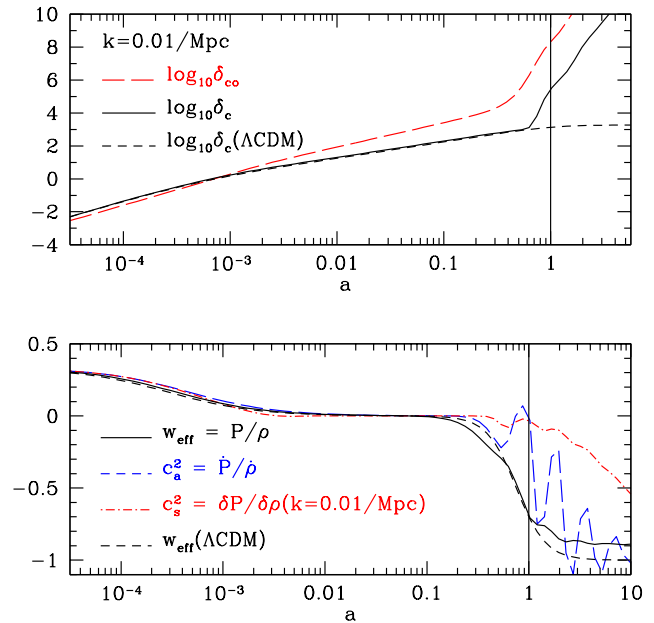}
\caption{[Bottom] The two component coupled dark energy (CDE)
  model, with $\lambda =2$ and coupling $C=-20$ with $H_0=70 \, {\rm km} \, {\rm s}^{-1}\, {\rm Mpc}^{-1}$, $\Omega_{b}=0.05$,
  $\Omega_{c}=0.2$, $\Omega_{co}=0.05$, and $\Omega_{V}=0.70$. At late times the scalar field finds the adiabatic minimum with asymptotic equation of state, and sound speed $= -1/(1+\gamma) = -0.89$, able to reproduce a viable background evolution consistent   with supernovae, CMB angular diameter distance and BBN expansion
  history constraints. The figure shows the evolution of the effective equation of state, $w_{eff} = P_{tot}/\rho_{tot}=(2/3) (d\ln t / d\ln a) -1,$ (black full line), adiabatic speed of sound, $c_{a}^{2}=\dot{P}_{tot}/\dot{\rho}_{tot}$, (blue long dashed line) and effective speed of sound for $c_s^2=\delta P_{tot}/\delta\rho_{tot}$ at $k=0.01/Mpc$ (red dot-dashed line).  The effective equation of state for a comparable $\Lambda$CDM model with $\Omega_{c}=0.25$, $\Omega_b=0.05$ and $\Omega_{\Lambda}=0.7$ is also shown (black dashed line). [Top] The growth of the fractional over-density $\delta=\delta\rho/\rho$ for $k=0.01/Mpc$ for the coupled CDM component, $\delta_{co}$, (red long dashed line) and uncoupled component, $\delta_{c}$, (black full line) in comparison to the growth for the $\Lambda$CDM model (black dashed line). At late times the adiabatic behavior triggers a dramatic increase in the rate of growth of both uncoupled and coupled components, leading to structure predictions inconsistent with observations.\label{fig1}}
\end{center}
\end{figure}

These expectations are confirmed~(figure~\ref{fig1}) by a numerical analysis
of a two component coupled model.
We use $\lambda=2$, $C=-20$, $H_0=70 \, {\rm km} \, {\rm s}^{-1}\,{\rm Mpc}^{-1}$,
baryon fractional energy density, $\Omega_{b}=0.05$, uncoupled CDM component, $\Omega_{c}=0.2$, coupled component, $\Omega_{co}=0.05$, and potential fractional energy density,
$\Omega_{V}=0.7$. We fix initial conditions of $\phi/\mpl=10^{-10}$ and
$\dot\phi=0$ at $a=10^{-10}$ (initial conditions at least within $\phi/\mpl=10^{-30}-1$ give the same evolution
because of a scalar dynamical attractor) and assume that the CDM components have the same initial
fractional density perturbations $\delta_c=\delta_{co}$, fixed by the
usual adiabatic initial conditions. As shown in the bottom panel of figure~\ref{fig1}, the
background evolution is consistent with a $\Lambda$CDM like
scenario, with $w_{\rm eff}=-0.69$ today, approaching
$w_{\rm eff} \sim -0.89$ asymptotically. In the top panel we see that, once the scalar field has entered the adiabatic regime, giving rise to accelerative expansion, the density perturbations undergo significantly increased growth, in stark contrast  to the $\Lambda$CDM  scenario in which accelerative expansion is typically associated with late-time suppression of growth.

In summary, these models provide a class of theories for which the
background cosmology is compatible with observations, but which are
ruled out by the adiabatic instability of the perturbations.

Another interesting class is the chameleon models~\cite{Khoury:2003aq,Khoury:2003rn}
for which the adiabatic regime
has been previously demonstrated in static solutions for macroscopic bodies like
the Earth, and also in cosmological models~\cite{Brax:2004qh}.
One well-studied example of these has
inverse power law potentials, together with the constant coupling function in~(\ref{constantcoupling}), for which the effective potential is then
\be
V_{\rm eff}(\phi,\rho_c) = \lambda M^4 \left( \frac{M }{\phi} \right)^n
 + e^{-\beta C \phi/\mpl} \rho_c \ ,
\label{neweffpot}
\ee
where $M$ is a mass scale and $n>0$ and $\lambda$ are constants.
The existence of a local minimum, and hence an adiabatic regime, in~(\ref{neweffpot}) requires
$C<0$.
We shall restrict attention to the regime $\rho_c \gg \rho_{\rm crit} \equiv n \lambda M^4 (-\beta C M/\mpl)^n$.
The sound speed squared is
\be
\frac{1}{c_s^2}  = -1 +  \frac{n+1}{\beta C} \frac{\mpl}{\phi} \ ,
\label{cs213}
\ee
which is always negative as expected.

The range of spatial scales ${\cal
  L}$ over which the instability operates for a given density $\rho_c \gg \rho_{\rm crit}$ is
  non-empty for $\beta |C| \gg 1$, and is given by
  \be
1 \ll \frac{ (n+1) (\beta C)^2 \rho_{\rm crit}}{\mpl^2} \left(
  \frac{\rho_c}{ \rho_{\rm crit} } \right)^{\frac{n+2}{n+1}}{\cal L}^2 \ll
\beta^2 C^2 \ .
\label{range2}
\ee
If $\phi$ behaves as dark energy, we require $\rho_{\rm crit} \sim H_0^2 \mpl^2$.  Then for $\rho_c \sim \rho_{\rm crit}$, the maximum lengthscale is of order $H_0^{-1}$, and the  minimum is $\sim (H_0 \beta |C|)^{-1}$.  Thus a large set of
cosmological models are in the unstable regime at $\rho_c \sim \rho_{\rm crit}$
(if $\beta |C| \gg 1$), ruling them out in this regime.

In this letter we have demonstrated the existence and broad applicability of the {\it adiabatic instability} - operating in models in which there exists a nontrivial coupling between dark matter and dark energy.
We have presented general expressions for the conditions under which the adiabatic instability is relevant, and, when so,
the lengthscales over which it operates. This work provides a new way to constrain interactions in the dark sector, and
heavily restricts the class of models consistent with cosmic acceleration.

In a companion paper~\cite{paper2}, we derive in detail the results presented in this letter, and apply the results to a wide class of
coupled models including couplings to both CDM and neutrinos.

\acknowledgments

We thank Ole Bjaelde,  Anthony Brookfield, Steen Hannestad, Carsten Van der Bruck, Ira Wasserman and Christoph Wetterich for useful discussions.
RB's work is supported by National Science Foundation grants AST-0607018 and PHY-0555216, EF's by
NSF grants PHY-0457200 and PHY-0555216, and MT's by NSF grant PHY-0354990 and by Research Corporation.


\begin{thebibliography}{99}

\bibitem{combined}
T.~Damour, G.~W.~Gibbons and C.~Gundlach,
Phys.\ Rev.\ Lett.\  {\bf 64}, 123 (1990);
S.~M.~Carroll,
Phys.\ Rev.\ Lett.\  {\bf 81}, 3067 (1998);
J.~P.~Uzan,
Phys.\ Rev.\  D {\bf 59}, 123510 (1999);
L.~Amendola,
Phys.\ Rev.\ D {\bf 62}, 043511 (2000);
R.~Bean and J.~Magueijo,
Phys.\ Lett.\ B {\bf 517}, 177 (2001);
R.~Bean,
Phys.\ Rev.\ D {\bf 64}, 123516 (2001);
E.~Majerotto, D.~Sapone and L.~Amendola,
arXiv:astro-ph/0410543;
R.~Fardon, A.~E.~Nelson and N.~Weiner,
 JCAP {\bf 0410}, 005 (2004);
{\it ibid},
JHEP {\bf 0603}, 042 (2006);
S.~Lee, G-C.~Liu and K-W.~Ng,
Phys. Rev. D {\bf 73}, 083516 (2006);
  S.~Tsujikawa,
  Phys.\ Rev.\  D {\bf 76}, 023514 (2007)
  [arXiv:0705.1032 [astro-ph]];
  N.~Agarwal and R.~Bean,
  arXiv:0708.3967 [astro-ph].

\bibitem{Das:2005yj}
  S.~Das, P.~S.~Corasaniti and J.~Khoury,
  Phys.\ Rev.\  D {\bf 73}, 083509 (2006)
  [arXiv:astro-ph/0510628].

\bibitem{Kesden:2006}
 M.~Kesden and M.~Kamionkowski,
arXiv:astro-ph/0608095.

\bibitem{Kaplinghat:2006jk}
  M.~Kaplinghat and A.~Rajaraman,
  arXiv:astro-ph/0601517.

\bibitem{Bjaelde:2007ki}
  O.~E.~Bjaelde, A.~W.~Brookfield, C.~van de Bruck, S.~Hannestad, D.~F.~Mota, L.~Schrempp and D.~Tocchini-Valentini,
  arXiv:0705.2018 [astro-ph].

\bibitem{Afshordi:2005ym}
 N.~Afshordi, M.~Zaldarriaga and K.~Kohri,
 Phys. Rev. D {\bf 72}, 065024 (2005).



\bibitem{combined1}
S.~Capozziello, S.~Carloni and A.~Troisi,
arXiv:astro-ph/0303041;
S.~M.~Carroll, V.~Duvvuri, M.~Trodden and M.~S.~Turner,
Phys.\ Rev.\ D {\bf 70}, 043528 (2004);
T. Chiba, Phys. Lett. B {\bf B575}, 1 (2003);
S.~M.~Carroll, I.~Sawicki, A.~Silvestri and M.~Trodden,
New J.\ Phys.\  {\bf 8}, 323 (2006);
L. Amendola, D. Polarski and S. Tsujikawa, Phys. Rev. Lett. {\bf 98}, 131302 (2007);
N. ~Agarwal and R. ~Bean,
arXiv:0708.3967.


\bibitem{paper2}
R.~Bean, E.~E.~Flanagan and M.~Trodden, 
arXiv:0709.1128.

\bibitem{Khoury:2003aq}
 J.~Khoury, A.~Weltman,
 Phys. Rev. Lett. {\bf 93}, 171190 (2004).

\bibitem{Khoury:2003rn}
J.~Khoury and A.~Weltman,
Phys. Rev. D {\bf 69}, 044026 (2004).

\bibitem{Brax:2004qh}
P.~Brax, C.~van de Bruck, A.C.~Davis, J.~Khoury and A.~Weltman
arXiv:astro-ph/0408415.

\bibitem{Sawicki:2007tf}
  I.~Sawicki and W.~Hu,
  arXiv:astro-ph/0702278.

\bibitem{Lewis:2002ah}
  A.~Lewis and S.~Bridle,
  Phys.\ Rev.\  D {\bf 66}, 103511 (2002)
  [arXiv:astro-ph/0205436].

\end{thebibliography}
\end{document}